# New Insights into the Dependence of Glass Transition Temperature and Dynamic Fragility on Molecular Weight in Oligomers and Polymers

Valeriy V. Ginzburg[1,*], Oleg V. Gendelman[2], Riccardo Casalini[3], and Alessio Zaccone[4]

[1]Department of Chemical Engineering and Materials Science, Michigan State University, East Lansing, Michigan, USA

[2]Faculty of Mechanical Engineering, Technion, Haifa, Israel

[3]Chemistry Division, Naval Research Laboratory, 4555 Overlook Avenue SW, Washington, D.C., USA

[4]University of Milan, Department of Physics, via Celoria 16, 20133 Milano, Italy

*Corresponding author, email ginzbur7@msu.edu

## Abstract

In an earlier preprint (V. V. Ginzburg, O. V. Gendelman, R. Casalini, and A. Zaccone, ArXiv preprint, https://arxiv.org/abs/2409.17291), we demonstrated that the "dynamic" (relaxation time) and "volume" equations of state for many amorphous polymers are "near-universal" – each material is characterized by two governing parameters – the glass transition temperature, $T_g$, and the elementary relaxation time, $\tau_{el}$. The dynamic fragility, $m$, is uniquely related to $\tau_{el}$. The earlier analysis was based on the data for high polymers (degree of polymerization, $N > 200$, and molecular weight, $M_w > 20,000$ g/mol). Here, we investigate the dependence of $T_g$ and $\tau_{el}$ on $N$ (or, equivalently, on the molecular weight, $M$). We consider the dielectric spectroscopy data for homologous series of four polymers: polystyrene (PS), poly-(2-vinylpyridine) (P2VP), poly(dimethylsiloxane) (PDMS), and polybutadiene (PB). It is shown that the relaxation time curves for various oligomers can be successfully collapsed onto the same master curve as the high polymers. The glass transition temperature and the dynamic fragility are found to be linear functions of $1/M$, in agreement with the Fox-Flory equation. The scaling results are interpreted based on the SL-TS2 theory, ensuring that the model description is consistent with the Boyer rules for the coefficients of thermal expansion above and below $T_g$, as well as the thermodynamical scaling for the pressure dependence of the relaxation time.

Thermodynamic and dynamic properties of amorphous polymers depend strongly on their molecular weight or degree of polymerization. In the 1950-s, Fox and Flory [1,2] investigated the dependence of the glass transition temperature, $T_g$, on the number-averaged molecular weight, $M_n$, of polystyrene (PS), and showed that it obeys a simple equation, $T_g(M_n) = T_{g,\infty} - K/M_n$, where $T_{g,\infty}$ is the high-polymer glass transition temperature and $K$ is a model parameter with units of *K kg mol$^{-1}$*. The physical meaning of this equation and the relationship between the parameter $K$ and the molecular characteristics of the polymer has been analyzed theoretically by Somcynsky and Patterson,[3] Sanmartin et al.,[4] and Zaccone and Terentjev.[5] Other empirical and first-principles equations (Ueberreiter-Kanig,[6] Gibbs-Dimarzio[7]) have also been proposed. Even so, there are many questions about the fundamental nature of the $T_g(M_n)$ dependence. As discussed by Drayer and Simmons,[8] the original rationale for the Fox-Flory and Ueberreiter-Kanig equations was that chain ends have different (higher) free volume and thus act as plasticizers, making the oligomers akin to miscible blends or random copolymers; more recent interpretations suggested the need to consider the impact of the chain ends on the rotational rigidity of the backbones. Drayer and Simmons simulated the dynamics of various chain types (a freely-joined chain, a moderately stiff chain, and a chemically realistic polymer molecule) and investigated the differences in the local dynamics of various monomers (close to the chain ends and close to the center of the molecule). They suggested that the combined effects of the chain end free volumes and the chain-end induced rigidity change could be described based on the Elastically Collective Nonlinear Langevin Equation (ECNLE) framework,[9–12] although the results are still preliminary and more analysis is needed. Baker *et al.* [13] re-analyzed the dynamic data on several glass-forming polymers and proposed a new theoretical framework in which the intrachain dihedral rotation barriers were assumed to be molecular-weight dependent; it was then shown that the dependence of $T_g$ on $M_n$ could be logarithmic for small $M_n$ and transitioning to Fox-

Flory behavior for large $M_n$; the transition molecular weight, $M^*$, was usually on the order of 0.5—2 kg/mol.

The above analysis suggests that oligomers are "more complex" than high polymers, in a sense that they are intermediate between the "small molecules" (where – as shown by Novikov and Rössler -- $T_g \propto \sqrt{M}$, $M$ being the molar mass of the glass-former)[14] and the "polymers" (where $T_g$ is independent of the molecular weight). Importantly, the above analysis was only devoted to the glass transition temperature (measured either in DSC or rheology). One could also consider the shape of the relaxation time vs. temperature curve and, in particular, the dynamic fragility, $m = \left[\dfrac{d\log(\tau)}{d(T_g/T)}\right]_{T=T_g}$, where $\tau$ is the α-relaxation time of the glass-former measured using dielectric spectroscopy or other techniques. The dependence of fragility on molecular weight in a number of polymers had been studied by multiple authors.[14–22] It has been suggested that there are at least three regions – low-molecular-weight, intermediate, and high-molecular-weight.[15] Hintermeyer et al.[15] associated the first crossover with the Rouse molecular weight, and the second crossover with the entanglement molecular weight, although other interpretations are also possible and no theoretical justification for this assignment exists.

Recently, Ginzburg and co-workers formulated a new two-state free-volume theory labeled SL-TS2 (Sanchez-Lacombe two-state, two-(time)scale model).[23–30] Within this approach, one can combine the "dynamic" (relaxation time vs. temperature and pressure) and "volume" (density vs. temperature and pressure) equations of state for glass-forming materials. In particular, it has been shown that for many polymers, the SL-TS2 parameters are "universal", meaning that each material is uniquely described by only two parameters (e.g., $T_g$ and fragility).[30] This universality was demonstrated for many different polymers in the high-molecular-weight limit. Here, we are extending the same analysis to oligomers and low-molecular-weight polymers and discuss how the universal dynamic and thermodynamic behavior is impacted by the finite molecular weight. In particular, we analyze the dielectric relaxation data for four sets of oligomers and polymers: polystyrene (PS),[20] polybutadiene (PB),[15] poly(dimethylsiloxane) (PDMS),[15] and poly(2-vinylpyridine) (P2VP).[22]

As a starting point, we demonstrate the scaling and universality according to the following transformation. The $\alpha$-relaxation time as a function of temperature is re-plotted as a "log-log" Arrhenius plot, and then the horizontal shift is applied to collapse all the curves onto one master curve. Furthermore, the master curve is then shifted vertically to overlap with the theoretical function described by,

$$\log\left(\frac{\tau_\alpha[T]}{\tau_{el}}\right) = \mathbb{F}[x] = (x^{-1} - 1)^{-\mu} \tag{1}$$

Here, the power-law exponent $\mu \approx 1.12$, $x = (T_x/T)$, and $T_x$ and $\tau_{el}$ are the model parameters related to the standard parameters $T_g$ and $m$ via,

$$T_g = T_x \left[1 + (2 - \log(\tau_{el}))\right]^{-1/\mu} \tag{2a}$$

$$m = \mu \left[1 + (2 - \log(\tau_{el}))\right]^{-1/\mu} \left[(2 - \log(\tau_{el}))\right]^{1+1/\mu} \tag{2b}$$

The above results have been derived in our earlier paper[30] by parameterizing the SL-TS2 theory to describe the "ideal Boyer polymer" (IBP), i.e., the polymers obeying the Boyer-Spencer[31,32] and Simha-Boyer[33] rules for thermal expansion. In the following, we assume that all the oligomers and polymers can be treated as IBP, as the simplest possible approximation.

In Figures 1 and 2, we show the original (Figure 1) and shifted (Figure 2) $\alpha$-relaxation time vs. temperature for PS, PB, P2VP, and PDMS with different molecular weights. The shifted curves are compared with the master equation (eq 1) and a good agreement is found.

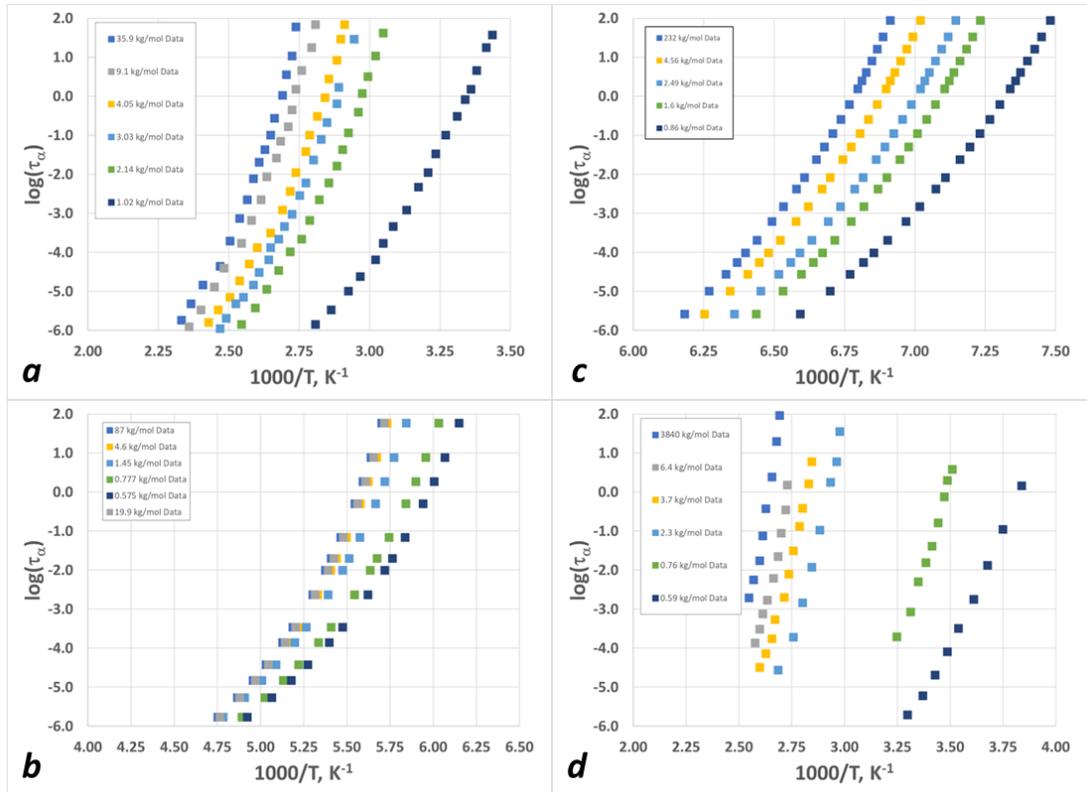

Figure 1. Dielectric α-relaxation time as function of inverse temperature for polymers with varying molecular weight. (a) P2VP; (b) PB; (c) PDMS; (d) PS.

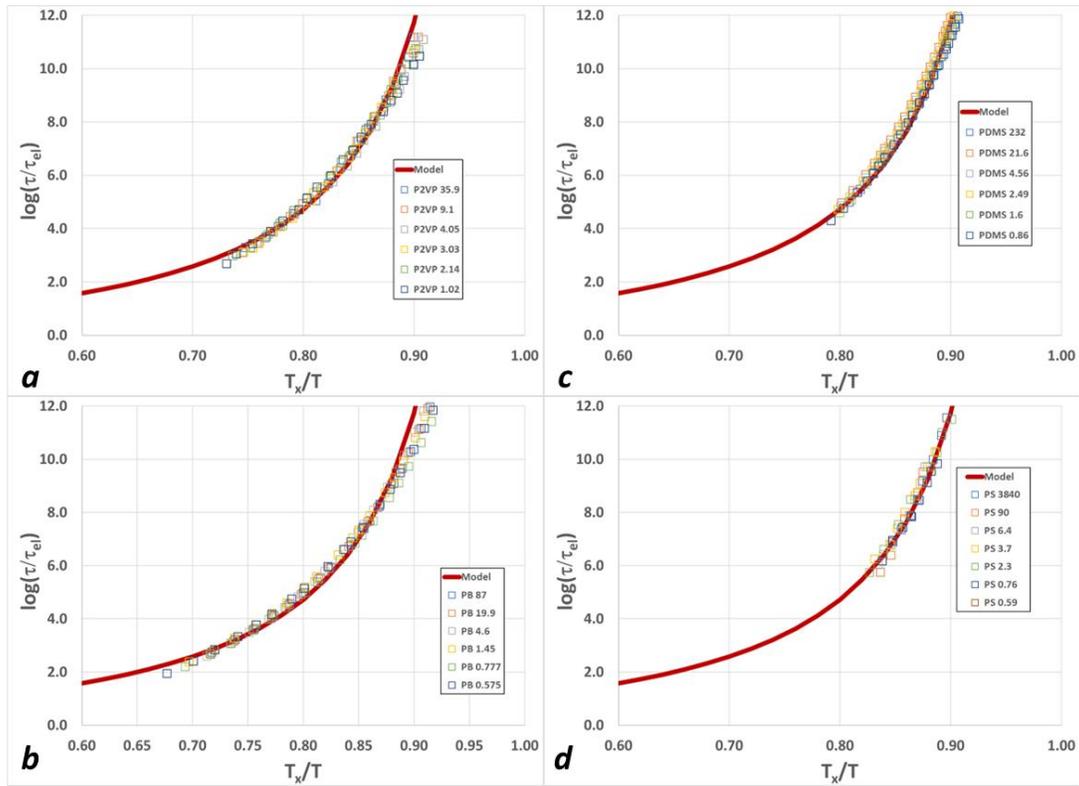

Figure 2. Scaled and shifted relaxation time plots and the master curve. (a) P2VP; (b) PB; (c) PDMS; (d) PS. The labels in the legend refer to molecular weight (in kg/mol). The red curve in all for panels is given by eq 1.

We can now investigate the dependence of the parameters $T_x$, $\tau_{el}$, $T_g$ (from eq 2a), and $m$ (from eq 2b) on molecular weight for each of the four sets. These dependencies are plotted in Figure 3. The dependence of the SL-TS2 thermodynamic transition, $T_x$, on the inverse molecular weight, $1/M$, is linear for all four polymer sets, consistent with the Fox-Flory equation (Figure 3a); same applies to the glass transition temperature, $T_g$ (Figure 3b). We considered only polymers and oligomers with M > 500 g/mol, so it is likely that our data do not capture the transition described by Baker et al.[13] The dependence of the fragility, $m$, on $1/M$ (Figure 3a) is also linear and shows no sign of any transition, although the relative standard error is relatively high (estimated at ±2.0 or about 2%).

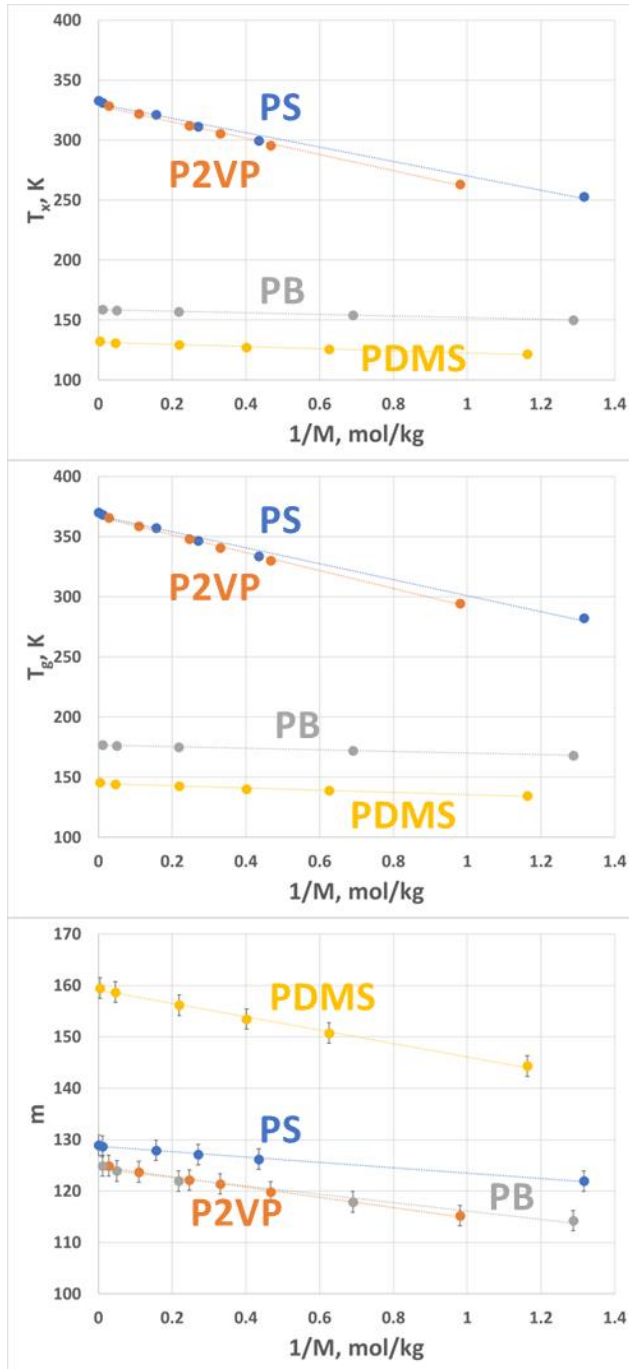

*Figure 3. Dependence of (a) SL-TS2 thermodynamic transition temperature; (b) glass transition temperature; (c) dynamic fragility on the inverse molecular weight. The lines indicate linear fits.*

The dependence of the model parameters on molecular weight is then given by,

$$T_X(M) = T_{X,\infty}\left[1 - \frac{M_X}{M}\right] \quad (3a)$$

$$T_g(M) = T_{g,\infty}\left[1 - \frac{M_g}{M}\right] \quad (3b)$$

$$m(M) = m_{\infty}\left[1 - \frac{M_m}{M}\right] \quad (3c)$$

The model parameters are summarized in Table 1.

Table 1. Model parameters for the four polymer sets

|  | P2VP | PB | PDMS | PS |
|---|---|---|---|---|
| $T_{X,\infty}$ | 329 | 159 | 131.5 | 330 |
| $M_X$, kg/mol | 0.206 | 0.044 | 0.066 | 0.179 |
| $T_{g,\infty}$ | 366.5 | 177 | 145 | 367 |
| $M_g$, kg/mol | 0.203 | 0.04 | 0.063 | 0.177 |
| $m_{\infty}$ | 125 | 124 | 159 | 129 |
| $M_m$, kg/mol | 0.081 | 0.067 | 0.082 | 0.04 |

It can be seen that the characteristic molecular weights, $M_X$, $M_g$, and $M_m$, are all very small (~40 – 200 g/mol), indicating that the standard Fox-Flory behavior is prevalent for all *M > 1000 g/mol* oligomers and high polymers. (The values of those parameters depend on the chemical nature of the oligomer/polymer, and at this time, we cannot speculate about these dependencies.) Within the SL-TS2 framework, it can be interpreted as an indication that the "IBP" assumption (thermal expansion according to the Boyer rules) is reasonable for all these systems. Certainly, this assumption does not hold for low-molecular-weight fluids,[34] so we expect the scaling and the Fox-Flory behavior that follows from it to fail.

In conclusion, we investigated the dependence of the relaxation time on temperature for several sets of glass-forming oligomers and polymers with varying molecular weights. We showed that all the curves can be superimposed on one master curve via simple horizontal and vertical shifts (using a log-log Arrhenius scale). The master curve is well-described by the "universal" SL-TS2 model calibrated for the "ideal Boyer polymers". The model parameters as functions of molecular weight are well-described by the Fox-Flory equation. The universality and scaling discussed here are expected to hold for high polymers and oligomers with (M > 1000 g/mol), while for monomers and short oligomers, different scaling rules are expected. Those will be the subject of future work.

## References

(1) Fox, T. G.; Flory, P. J. The Glass Temperature and Related Properties of Polystyrene. Influence of Molecular Weight. *Journal of Polymer Science* **1954**, *14* (75), 315–319.

(2) Fox Jr, T. G.; Flory, P. J. Second-order Transition Temperatures and Related Properties of Polystyrene. I. Influence of Molecular Weight. *J Appl Phys* **1950**, *21* (6), 581–591.

(3) Somcynsky, T.; Patterson, D. The Glass Transition and the Reduced Temperature of Polymeric Liquids. *Journal of Polymer Science* **1962**, *62* (174), S151–S155.

(4) Pezzin, G.; Zilio-Grandi, F.; Sanmartin, P. The Dependence of the Glass Transition Temperature on Molecular Weight for Polyvinylchloride. *Eur Polym J* **1970**, *6* (7), 1053–1061. https://doi.org/https://doi.org/10.1016/0014-3057(70)90038-8.

(5) Zaccone, A.; Terentjev, E. M. Disorder-Assisted Melting and the Glass Transition in Amorphous Solids. *Phys Rev Lett* **2013**, *110* (17), 178002.

(6) Ueberreiter, K.; Kanig, G. Self-Plasticization of Polymers. *J Colloid Sci* **1952**, *7* (6), 569–583. https://doi.org/https://doi.org/10.1016/0095-8522(52)90040-8.

(7) Gibbs, J. H.; DiMarzio, E. A. Nature of the Glass Transition and the Glassy State. *J Chem Phys* **1958**, *28* (3), 373–383.

(8) Drayer, W. F.; Simmons, D. S. Is the Molecular Weight Dependence of the Glass Transition Temperature Driven by a Chain End Effect? *Macromolecules* **2024**, *57* (12), 5589–5597. https://doi.org/10.1021/acs.macromol.4c00419.

(9) Phan, A. D.; Schweizer, K. S. Elastically Collective Nonlinear Langevin Equation Theory of Glass-Forming Liquids: Transient Localization, Thermodynamic Mapping, and Cooperativity. *Journal of Physical Chemistry B* **2018**, *122* (35), 8451–8461. https://doi.org/10.1021/acs.jpcb.8b04975.

(10) Mirigian, S.; Schweizer, K. S. Elastically Cooperative Activated Barrier Hopping Theory of Relaxation in Viscous Fluids. I. General Formulation and Application to Hard Sphere Fluids. *J Chem Phys* **2014**, *140* (19), 194506.

(11) Mirigian, S.; Schweizer, K. S. Elastically Cooperative Activated Barrier Hopping Theory of Relaxation in Viscous Fluids. II. Thermal Liquids. *J Chem Phys* **2014**, *140* (19), 194507.

(12) Zhou, Y.; Mei, B.; Schweizer, K. S. Activated Relaxation in Supercooled Monodisperse Atomic and Polymeric WCA Fluids: Simulation and ECNLE Theory. *J. Chem. Phys.* **2022**, *156*, 114901.

(13) Baker, D. L.; Reynolds, M.; Masurel, R.; Olmsted, P. D.; Mattsson, J. Cooperative Intramolecular Dynamics Control the Chain-Length-Dependent Glass Transition in Polymers. *Phys Rev X* **2022**, *12* (2), 21047. https://doi.org/10.1103/PhysRevX.12.021047.

(14) Novikov, V.; Rössler, E. Correlation between Glass Transition Temperature and Molecular Mass in Non-Polymeric and Polymer Glass Formers. *Polymer (Guildf)* **2013**, *54*, 6987.

(15) Hintermeyer, J.; Herrmann, A.; Kahlau, R.; Goiceanu, C.; Rössler, E. A. Molecular Weight Dependence of Glassy Dynamics in Linear Polymers Revisited. *Macromolecules* **2008**, *41* (23), 9335–9344. https://doi.org/10.1021/ma8016794.

(16) Roland, C. M.; Ngai, K. L.; Plazek, D. J. Modes of Molecular Motion in Low Molecular Weight Polystyrene. *Macromolecules* **2004**, *37*, 7051.

(17) Ding, Y.; Kisliuk, A.; Sokolov, A. P. When Does a Molecule Become a Polymer? *Macromolecules* **2004**, *37*, 161.

(18) Rizos, A. K.; Ngai, K. L. Local Segmental Dynamics of Low Molecular Weight Polystyrene: New Results and Interpretation. *Macromolecules* **1998**, *31*, 6217.

(19) Santangelo, P. G.; Roland, C. M. Molecular Weight Dependence of Fragility in Polystyrene. *Macromolecules* **1998**, *31* (14), 4581–4585. https://doi.org/10.1021/ma971823k.

(20) Roland, C. M.; Casalini, R. Temperature Dependence of Local Segmental Motion in Polystyrene and Its Variation with Molecular Weight. *J. Chem. Phys.* **2003**, *119*, 1838.

(21) Agapov, A. L.; Sokolov, A. P. Does the Molecular Weight Dependence of $T_g$ Correlate to $M_e$? *Macromolecules* **2009**, *42* (7), 2877–2878.

(22) Agapov, A. L.; Novikov, V. N.; Hong, T.; Fan, F.; Sokolov, A. P. Surprising Temperature Scaling of Viscoelastic Properties in Polymers. *Macromolecules* **2018**, *51* (13), 4874–4881. https://doi.org/10.1021/acs.macromol.8b00454.

(23) Ginzburg, V. A Simple Mean-Field Model of Glassy Dynamics and Glass Transition. *Soft Matter* **2020**, *16* (3), 810–825. https://doi.org/10.1039/c9sm01575b.

(24) Ginzburg, V. Combined Description of Polymer PVT and Relaxation Data Using a Dynamic "SL-TS2" Mean-Field Lattice Model. *Soft Matter* **2021**, *17*, 9094–9106. https://doi.org/10.1039/D1SM00953B.

(25) Ginzburg, V. Modeling the Glass Transition and Glassy Dynamics of Random Copolymers Using the TS2 Mean-Field Approach. *Macromolecules* **2021**, *54* (6), 2774–2782. https://doi.org/10.1021/acs.macromol.1c00024.

(26) Ginzburg, V. Modeling the Glass Transition of Free-Standing Polymer Thin Films Using the "SL-TS2" Mean-Field Approach. *Macromolecules* **2022**, *55* (3), 873–882. https://doi.org/10.1021/acs.macromol.1c02370.

(27) Ginzburg, V. V; Zaccone, A.; Casalini, R. Combined Description of Pressure-Volume-Temperature and Dielectric Relaxation of Several Polymeric and Low-Molecular-Weight Organic Glass-Formers Using 'SL-TS2' Mean-Field Approach. *Soft Matter* **2022**, *18*, 8456–8466.

(28) Ginzburg, V. V; Fazio, E.; Corsaro, C. Combined Description of the Equation of State and Diffusion Coefficient of Liquid Water Using a Two-State Sanchez–Lacombe Approach. *Molecules* **2023**, *28* (6), 2560.

(29) Ginzburg, V. V; Gendelman, O. V; Zaccone, A. Unifying Physical Framework for Stretched-Exponential, Compressed-Exponential, and Logarithmic Relaxation Phenomena in Glassy Polymers. *Macromolecules* **2024**. https://doi.org/10.1021/acs.macromol.3c02480.

(30) Ginzburg, V. V; Gendelman, O. V; Casalini, R.; Zaccone, A. Universality of Dynamic and Thermodynamic Behavior of Polymers near Their Glass Transition. *arXiv preprint arXiv:2409.17291* **2024**.

(31)　Boyer, R. F.; Spencer, R. S. Thermal Expansion and Second-order Transition Effects in High Polymers: Part I. Experimental Results. *J Appl Phys* **1944**, *15* (4), 398–405.

(32)　Boyer, R. F.; Spencer, R. S. Thermal Expansion and Second-Order Transition Effects in High Polymers: PART II. Theory. *J Appl Phys* **1945**, *16* (10), 594–607. https://doi.org/10.1063/1.1707509.

(33)　Simha, R.; Boyer, R. F. On a General Relation Involving the Glass Temperature and Coefficients of Expansion of Polymers. *J Chem Phys* **1962**, *37* (5), 1003–1007.

(34)　Lunkenheimer, P.; Loidl, A.; Riechers, B.; Zaccone, A.; Samwer, K. Thermal Expansion and the Glass Transition. *Nat Phys* **2023**, *19* (5), 694–699. https://doi.org/10.1038/s41567-022-01920-5.